\documentclass[a4paper,10pt]{article}
\usepackage[top=1.0in,right=1.in,left=1.in,bottom=1.4in]{geometry}
\usepackage{amssymb,amsbsy}
\usepackage{amsmath}
\usepackage{amsthm}
\usepackage{amsfonts}
\usepackage{graphics,graphicx}
\usepackage[T1]{fontenc}
\usepackage{color}
\usepackage{booktabs}
\usepackage[affil-it,auth-sc]{authblk}
\usepackage{subfigure}

\newcommand{\beq}{\begin{equation}}
\newcommand{\eeq}{\end{equation}}

\newcommand{\beqar}{\begin{eqnarray}}
\newcommand{\eeqar}{\end{eqnarray}}
\newcommand{\bit}{\begin{itemize}}
\newcommand{\eit}{\end{itemize}}
\newcommand{\benum}{\begin{enumerate}}
\newcommand{\eenum}{\end{enumerate}}
\newcommand{\barr}{\begin{array}}
\newcommand{\earr}{\end{array}}
\def\ds{\displaystyle}

\newcommand{\modIII}{\text{III}}

\def\XXint#1#2#3{{\setbox0=\hbox{$#1{#2#3}{\int}$}
   \vcenter{\hbox{$#2#3$}}\kern-.5\wd0}}

\def\b0{\mbox{\boldmath $0$}}

\def\bc{\mbox{\boldmath $c$}}

\def\bY{\mbox{\boldmath $Y$}}

\def\f0{\ensuremath{\mathbb{O}}}

\newcommand{\bmM}{\mbox{\boldmath $\mathcal{M}$}}

\title{Mode III crack propagation in a bimaterial plane driven by a channel of small line defects}

\author[1]{A. Piccolroaz\footnote{Corresponding author: e-mail: roaz@ing.unitn.it; phone: +39\,0461\,282583.}}
\author[1]{G. Mishuris}
\author[2]{A. Movchan}
\author[2]{N. Movchan}

\affil[1]{Institute of Mathematical and Physical Sciences, Aberystwyth University, Wales, UK}
\affil[2]{Department of Mathematical Sciences, University of Liverpool, Liverpool, L69 3BX, UK}

\begin{document}

\maketitle

\begin{abstract}

We consider the quasi-static propagation of a Mode III crack along the interface in a bimaterial plane containing a finite array of small line defects 
(microcracks and rigid line inclusions). The microdefects are arranged to form a channel around the interface that can facilitate (or prevent) the crack propagation. 
The two dissimilar elastic materials are assumed to be weakly bonded, so that there is no kinking of the main crack from the straight path. On the basis of 
asymptotic formulae obtained by the authors, the propagation is analysed as a perturbation problem and the incremental crack advance is analytically derived at 
each position of the crack tip along the interface relative to the position of the defects. Numerical examples are provided showing potential applications of the 
proposed approach in the analysis of failure of composite materials. Extension to the case of infinite number of defects is discussed.

\vspace{3mm}

\end{abstract}

% \tableofcontents

% \newpage

\section{Introduction}
\label{intro}

There is a vast literature devoted to the problem of the interaction between a stationary main crack and microdefects, in particular microcracks. The methods of 
solution range from complex potentials and the theory of analytic functions (\cite{gong-1989,gong-1992,gong-1995}), singular integral equations 
(\cite{romalis-1984,rubinstein-1986,rubinstein-1988,tamuzs-1993,tamuzs-1996}), and the weight function and dipole matrix approach (\cite{bigoni-1998,movchan-2002,
roaz-commat2011}). However, the analysis of crack propagation, even in the quasi-static case, requires also the analysis of singular perturbations produced by a 
small advance of the crack tip across the microdefect field (\cite{roaz-jmps2009,roaz-ijss2011}).

The symmetric and skew-symmetric weight functions for a two-dimensional interfacial crack, recently derived by \cite{roaz-jmps2009}, can be efficiently 
employed to analyse the propagation of cracks along the interface in heterogeneous materials with small defects. In particular, the notion of skew-symmetric function 
is essential for the problem under consideration , since the ``effective'' loading produced by the defects on the crack faces is not symmetrical, in general. 

We consider the quasi-static propagation of a Mode III crack along the interface in a bimaterial plane containing a finite array of small line defects. The microdefects 
are arranged to form a channel around the interface that can facilitate (or prevent) the crack propagation. The defect in question are microcracks and rigid line inclusion. 
The two elastic half-planes are assumed to be weakly bonded, so that there is no kinking of the main crack from the straight path. The defects are small compared to the 
distance from the crack tip and the distance between each other and, consequently, the interaction between them can be neglected. Using the superposition principle, each 
defect can be analysed separately and then the solution found by summing the contributions for all defects. 

\begin{figure}[!htcb]
\begin{center}
\includegraphics[width=6cm]{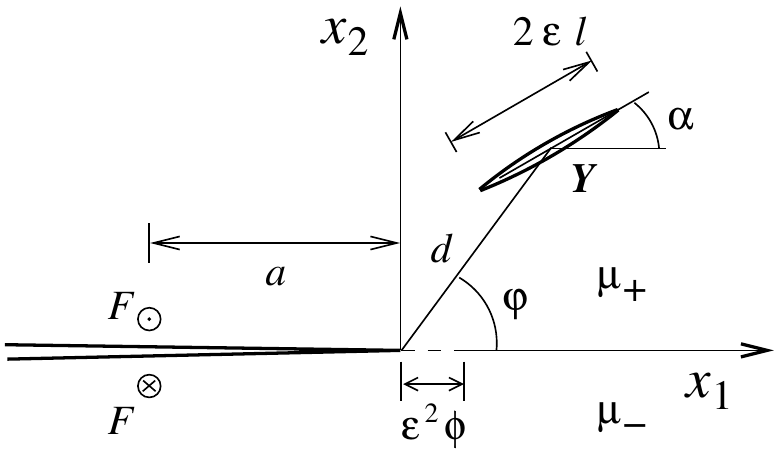}
\caption{\footnotesize A small line defect interacting with an interfacial crack in a bimaterial plane.}
\label{fig01}
\end{center}
\end{figure}

Fig.\ \ref{fig01} shows a main crack at the interface between two dissimilar elastic material interacting with a small line defect at a finite distance from 
the crack tip. The shear moduli for the upper and lower half-planes are denoted by $\mu_+$ and $\mu_-$, respectively. The line defect, which can be a microcrack or a 
movable rigid line inclusion, has length $2 \varepsilon l$, where $\varepsilon > 0$ is a small dimensionless parameter, and its orientation with respect to the 
$x_1$-axis is determined by the angle $\alpha$. The centre of the defect is denoted by $\bY$, having polar coordinates $d$ and $\varphi$. The structure is loaded at 
the crack surfaces by two symmetrical point forces placed at a distance $a$ from the crack tip. The crack advance is denoted by $\varepsilon^2 \phi$.
%%%%%%%%%%%%%%%%%%%%%%%%%%%%%%%%%%%%%%%%%%%%%%%%%%%%%%%%%%%%%%%%%%%%%%
%%%%%%%%%%%%%%%%%%%%%%%%%%%%%%%%%%%%%%%%%%%%%%%%%%%%%%%%%%%%%%%%%%%%%%

The detailed asymptotic analysis for this problem can be found in \cite{roaz-ijss2011}, where the approach based on weight function and dipole matrix is used. 
In particular, the dipole fields for different types of defect are constructed and used together with symmetric and skew-symmetric weight functions for an interfacial 
crack (\cite{roaz-jmps2009}) to derive the corresponding perturbation of the stress intensity factor acting at the main crack tip.

In the next section, we summarise the asymptotic formulae needed for the analysis of propagation of the crack in a finite channel of defects, which is presented in 
Section~\ref{secfin}.

\section{Asymptotic formulae}
\label{secasy}

The propagation of the main crack is described as a perturbation problem, in which the stress intensity factor is expanded as follows
\beq
\label{expan}
K_\modIII = K_\modIII^{(0)} + \varepsilon^2 \left(\Delta K_\modIII^\phi + 
\sum_{j=1}^{N} \Delta K_\modIII^{(j)}\right) + o(\varepsilon^2), \quad \varepsilon \to 0,
\eeq
where $K_\modIII^{(0)} = \sqrt{2 / \pi a}\, F$ is the stress intensity factor for the unperturbed crack, $\Delta K_\modIII^\phi$ is the singular perturbation due to 
a small advance $\varepsilon^2 \phi$ of the crack tip along the interface, and $\sum_{j=1}^{N} \Delta K_\modIII^{(j)}$ is the perturbation produced by a finite 
number $N$ of defects.

Following \cite{roaz-ijss2011}, the asymptotic formulae for the perturbation terms in (\ref{expan}) are given by
\beq\nonumber
\Delta K_\modIII^\phi = \frac{\phi}{2} A_\modIII^{(0)}, \quad 
\Delta K_\modIII^{(j)} = - \sqrt{\frac{2}{\pi}} \frac{\mu_+\mu_-}{\mu_+ + \mu_-} \nabla u^{(0)}(\bY_j) \cdot \bmM_j \bc_j,
\eeq
\beq
\bc_j = \frac{1}{2d_j^{3/2}}\left[-\sin\frac{3\varphi_j}{2},\cos\frac{3\varphi_j}{2}\right],
\eeq
where $A_\modIII^{(0)} = -\sqrt{2 / \pi a}\, F/a$ is the coefficient in the second-order term asymptotics of the unperturbed solution, $\nabla u^{(0)}(\bY_j)$ is 
the gradient of the unperturbed solution evaluated at $\bY_j$, the centre of the defect, and $\bmM_j$ is the dipole matrix associated with the type of defect:

\noindent
-- for a microcrack: 
\beq 
\bmM_j = -\frac{\pi l_j^2}{2} \left[
\barr{cc}
1 - \cos 2\alpha_j & -\sin 2\alpha_j \\[3mm]
-\sin 2\alpha_j & 1 + \cos 2\alpha_j
\earr
\right],
\eeq
-- for a movable rigid line inclusion :
\beq 
\bmM_j = \frac{\pi l_j^2}{2} \left[
\barr{cc}
1 + \cos 2\alpha_j & \sin 2\alpha_j \\[3mm]
\sin 2\alpha_j & 1 - \cos 2\alpha_j 
\earr 
\right]. 
\eeq 
The full-field solution $u_0$ for a Mode III crack in a bimaterial plane without defects can be found in \cite{roaz-ijf2010,roaz-commat2011}.

Assuming that the crack is propagating quasi-statically along the interface, the stress intensity factor remains constant and equal to the critical value, so that 
the perturbation term in parenthesis in (\ref{expan}) equals zero and we get the formula
\beq
\label{phi}
\phi = -\frac{2}{A_\modIII^{(0)}} \sum_{j=1}^{N} \Delta K_\modIII^{(j)},
\eeq
which gives the incremental crack advance at each position of the crack tip, driven by the presence of the defects, $\phi > 0$. The procedure is iterated and the total 
crack elongation at arrest (corresponding to the first occurrence of the condition $\phi \le 0$) is computed as
\beq
x(K) = \varepsilon^2 \sum_{i=0}^{K} \phi_i,
\eeq
where $K$ is the number of iterations.

\section{Numerical computations}
\label{secfin}

We consider the propagation of the main crack driven by a finite number of small line defects arranged to form a channel around the interface. Two different examples 
are discussed below, one consisting of only microcracks and the second consisting of microcracks and rigid line inclusions.

\begin{figure}[!htcb]
\begin{center}
\subfigure[microcracks]{\includegraphics[width=6cm]{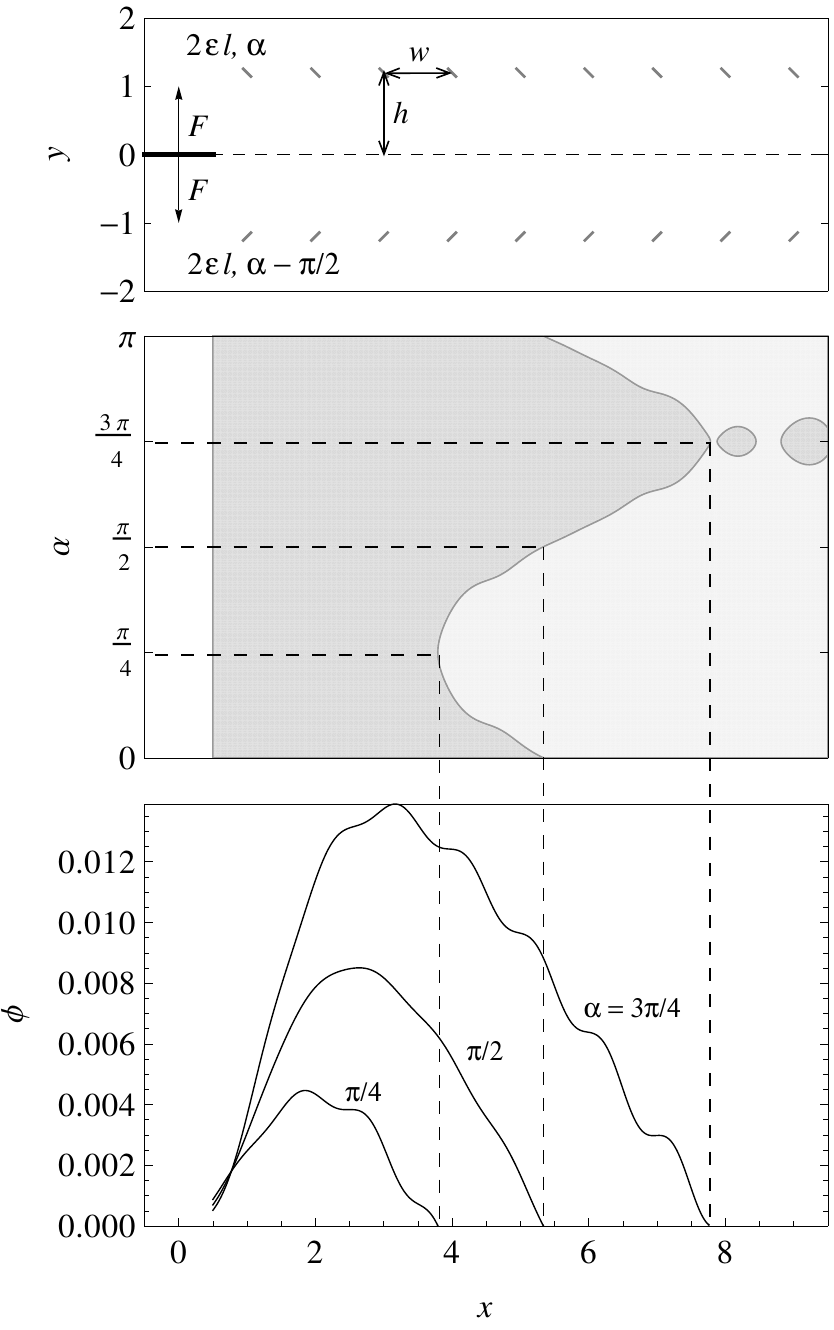}}\qquad
\subfigure[rigid line inclusions/microcracks]{\includegraphics[width=6cm]{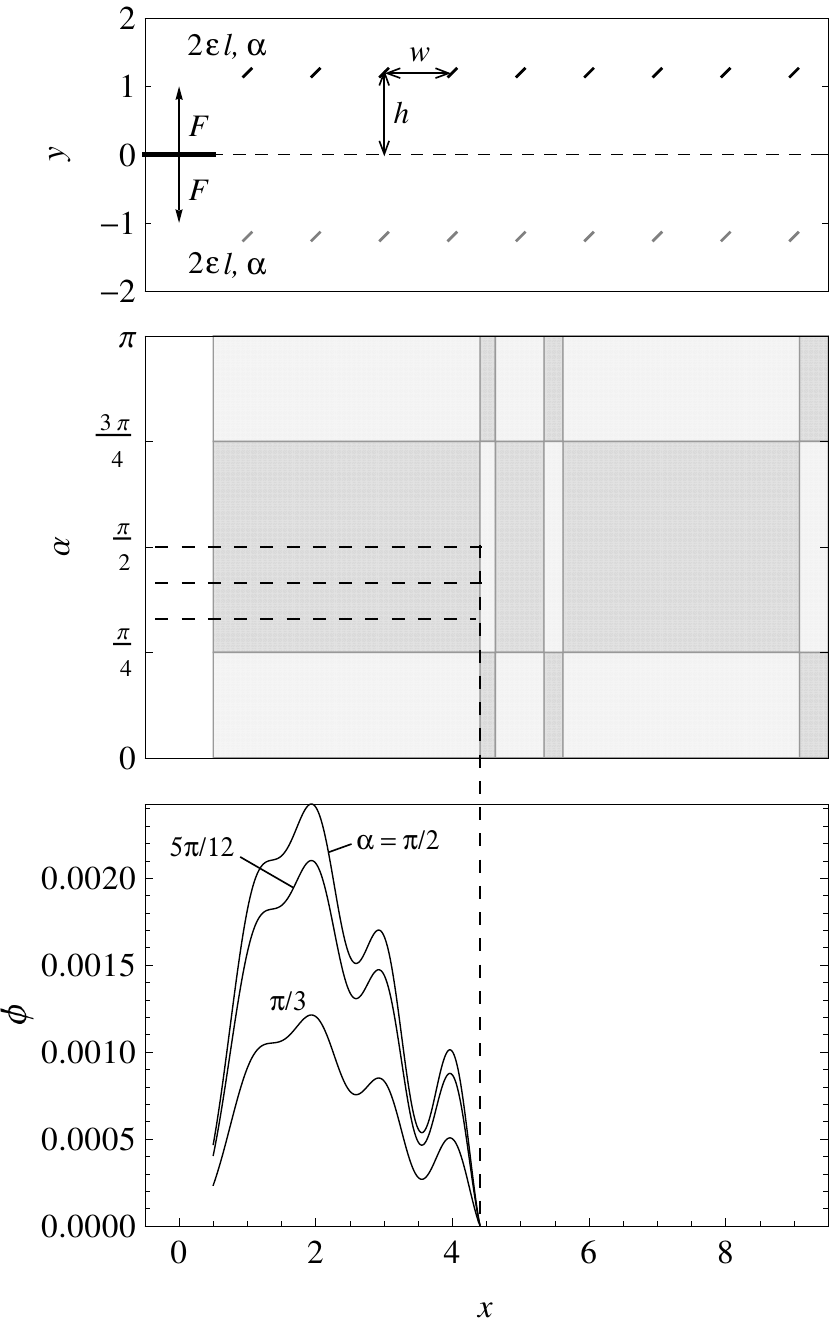}}
\caption{\footnotesize Upper part: (a) Main crack propagating along a weak interface between identical materials containing a 2$\times$9 array of microcracks. 
(b) Main crack propagating along a weak interface between identical materials containing a 2$\times$9 array consisting of nine rigid line 
inclusions in the upper half-plane and nine microcracks in the lower half-plane. Middle part: Shielding-amplification diagram in the $\alpha$ vs. $x$ plane. 
Lower part: Incremental crack advance $\phi$ as a function of the crack tip position $x$ for three values of $\alpha$.}
\label{figchA2B2n1}
\end{center}
\end{figure}

The first example is shown in Fig.\ \ref{figchA2B2n1}a, upper part, consisting in 18 microcracks arranged in a 2 x 9 array. All the microcracks have the same length, 
$2\varepsilon l = 0.2$, and distance from the interface, $h = 1.2$, measured from the centre of the microcracks. The parameter $\alpha$ denotes the inclination 
of the microcracks in the first row, while microcracks in the second row have inclination $\alpha - \pi/2$, so that microcracks in the lower half-plane are 
perpendicular to the microcracks in the upper half-plane. The microcracks, above and below the interface, are separated one from the another by a constant distance, 
$w = 1$. The crack is loaded by a two-point symmetrical force applied at a distance $a = 0.5$ from the crack tip. We analyse first the case of a weak interface in a  
homogeneous plane, that is the contrast parameter $\eta = (\mu_- - \mu_+)/(\mu_- + \mu_+)$ is set equal to zero.
%%%%%%%%%%%%%%%%%%%%%%%%%%%%%%%%%%%%%%%%%%%%%%%%%%%%%%%%%%%%%%%%%%%%%%
%%%%%%%%%%%%%%%%%%%%%%%%%%%%%%%%%%%%%%%%%%%%%%%%%%%%%%%%%%%%%%%%%%%%%%

In the middle part of the figure we show the corresponding shielding-amplification diagram. We have shielding effect when the defects produce a decrease of the 
stress intensity factor, $\sum_j \Delta K_\modIII^{(j)}/K_\modIII^{(0)} < 0$, or amplification effect in the opposite case, 
$\sum_j \Delta K_\modIII^{(j)}/K_\modIII^{(0)} > 0$ (\cite{ortiz-1987,hutchinson-1987,loehnert-2007}).
On the horizontal axis we specify the position $x$ of the crack tip along the interface, measured from the point of application of the concentrated forces, while the 
vertical axis stands for the angle $\alpha$ ($0 < \alpha < \pi$). 
For three values of $\alpha$, namely $\alpha = \pi/4$, $\pi/2$, $3\pi/4$, we report in the lower part of the figure the incremental crack advance $\phi$ as a function 
of the crack tip position $x$. In the initial position, $x = 0.5$, the perturbation of the stress intensity factor is positive (amplification) and consequently the 
crack propagates, driven by the defects, and it stops when a neutral configuration is reached (line of separation between amplification and shielding regions).
It appears that it is not possible to propagate the crack up to the end of the channel. Results show that the value $\alpha = 3\pi/4$ is the most favourable for the 
crack propagation.

The geometric configuration for the second example is shown in Fig.\ \ref{figchA2B2n1}b, upper part. It consists of a row of nine rigid line inclusions in the 
upper half-plane and a row of nine microcracks in the lower half-plane. All defects have the same length $2\varepsilon l = 0.2$, inclination $\alpha$ and 
distance from the interface $h = 1.2$ and they are horizontally separated one from the another by the same distance $w = 1$. The main crack is loaded by a two-point 
force applied at a distance $a = 0.5$ from the crack tip.  In this case, the shielding-amplification diagram shows no dependence on the orientation of the defects, 
angle $\alpha$. Consequently, the total crack elongation before arrest is independent of $\alpha$ and the crack stops approximately in the middle of the array, 
see the three propagation curves in the lower part of the figure.

\begin{figure}[!htcb]
\begin{center}
\subfigure[microcracks]{\includegraphics[width=6cm]{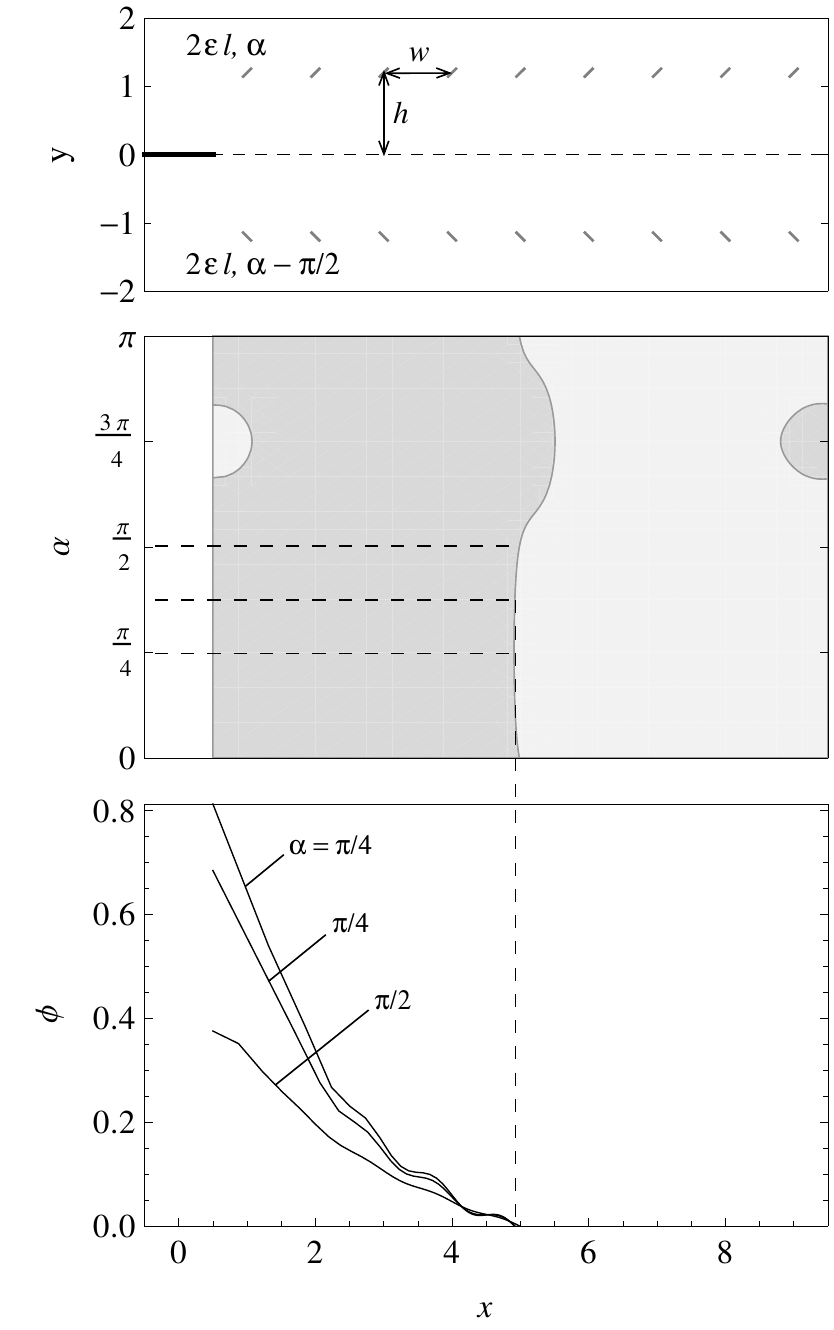}}\qquad 
\subfigure[rigid line inclusions/microcracks]{\includegraphics[width=6cm]{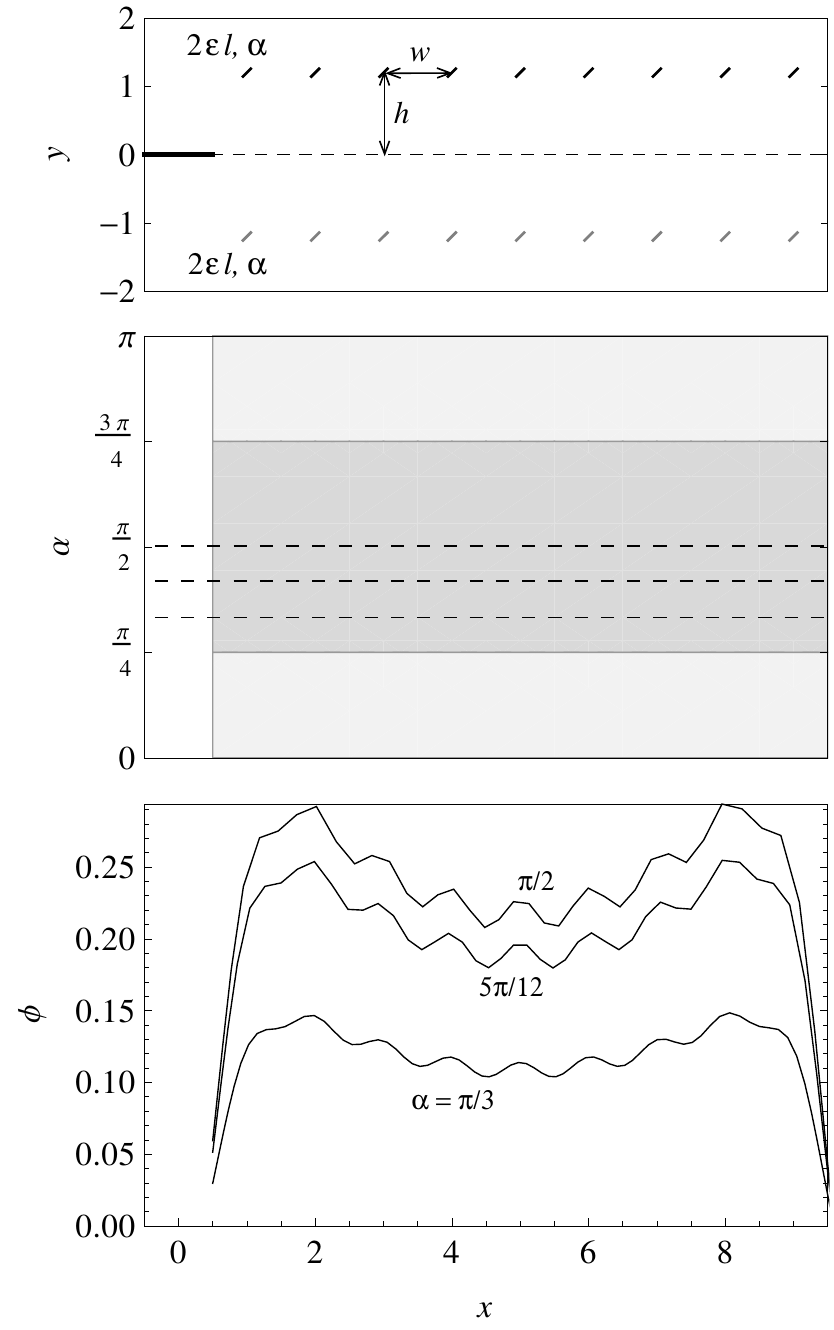}}
\caption{\footnotesize Loading applied at large distance from the crack tip. Figure (a) corresponds to the case shown in Fig.\ \ref{figchA2B2n1}a but with two-point 
forces applied at distance $a = 100$ from the crack tip. Figure (b) corresponds to the case shown in Fig.\ \ref{figchA2B2n1}b but with two-point forces applied at 
distance $a = 100$ from the crack tip.} 
\label{figchA2B2n1w1=100}
\end{center}
\end{figure}

The effect of the position of the two-point force with respect to the crack tip is illustrated in Fig.\ \ref{figchA2B2n1w1=100}. Fig.\ \ref{figchA2B2n1w1=100}a 
corresponds to the case shown in Fig.\ \ref{figchA2B2n1}a but with two-point force applied at distance $a = 100$ from the crack tip, instead of $a = 0.5$. 
Fig.\ \ref{figchA2B2n1w1=100}b corresponds to the case shown in Fig.\ \ref{figchA2B2n1}b but with two-point force applied at distance $a = 100$ from the crack tip. 
%%%%%%%%%%%%%%%%%%%%%%%%%%%%%%%%%%%%%%%%%%%%%%%%%%%%%%%%%%%%%%%%%%%%%%
%%%%%%%%%%%%%%%%%%%%%%%%%%%%%%%%%%%%%%%%%%%%%%%%%%%%%%%%%%%%%%%%%%%%%%

Two interesting observations can be made. First, we may note that the ``speed'' of the crack advance is much higher compared to the case when the force is applied 
close to the crack tip (the incremental crack advance $\phi$ is higher by two orders). This can be explained by noticing that, from the unperturbed solution $u_0$ 
reported in \cite{roaz-commat2011}, we deduce
\beq
\nabla u^{(0)} = O(a^{-1/2}), \quad A_\modIII^{(0)} = O(a^{-3/2}), \quad a \to \infty,
\eeq
and consequently
\beq
\phi = O(a), \quad a \to \infty.
\eeq
Second, the crack propagates up to the centre of the channel, $x = 5$, in the first case, see Fig.\ \ref{figchA2B2n1w1=100}a, while it propagates across and beyond 
the whole array in the second case, see Fig.\ \ref{figchA2B2n1w1=100}b. Furthermore, the propagation curves in Fig.\ \ref{figchA2B2n1w1=100}b show perfect symmetry 
with respect to the centre of the array. 

\begin{figure}[!htcb]
\begin{center}
\includegraphics[width=6cm]{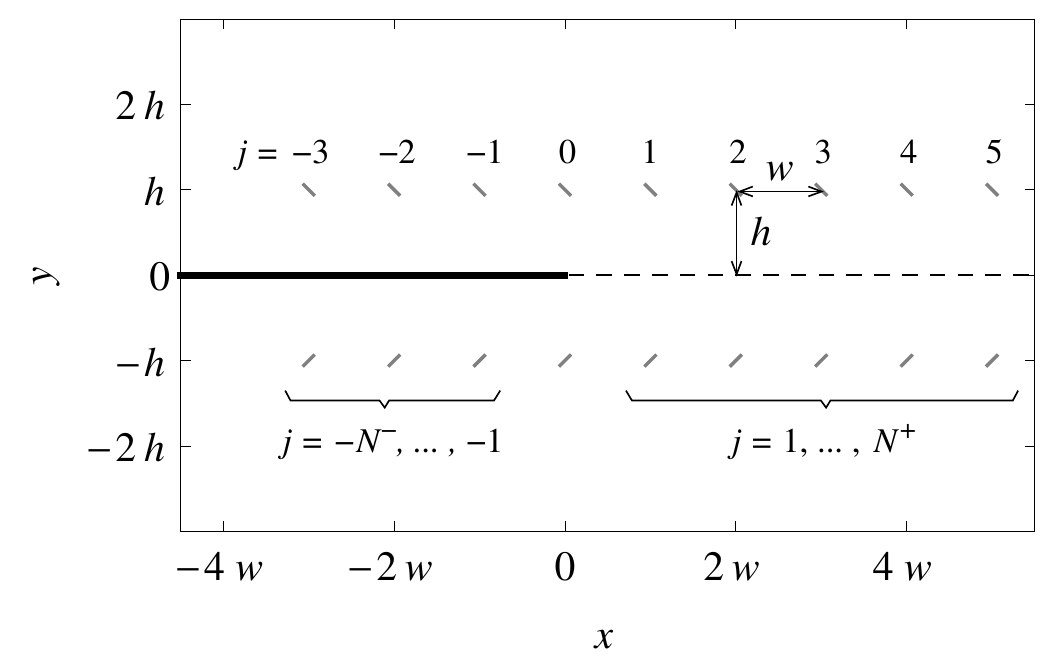}
\caption{\footnotesize Finite array of defects horizontally separated by a length $w$ from each other and by a length $h$ from the interface.} 
\label{figsumA2n1j}
\end{center}
\end{figure}

All these observations can be confirmed analytically by deriving the asymptotic approximation of $\phi$ as $a \to \infty$. In particular, we assume the configuration 
shown in Fig.\ \ref{figsumA2n1j}. The crack tip is situated in correspondence of the line joining two defects on opposite sides of the interface. Moreover, we 
number the defects with index $j$ running from 1 to $N^+$ for the defects ahead of the crack tip, and from $-N^-$ to $-1$ for the defects behind the crack tip. 
The defects right above and below the crack tip are numbered $j = 0$.
%%%%%%%%%%%%%%%%%%%%%%%%%%%%%%%%%%%%%%%%%%%%%%%%%%%%%%%%%%%%%%%%%%%%%%
%%%%%%%%%%%%%%%%%%%%%%%%%%%%%%%%%%%%%%%%%%%%%%%%%%%%%%%%%%%%%%%%%%%%%%

For the case of Fig.\ \ref{figchA2B2n1w1=100}a, microcracks above and below the interface, we use the formula valid for a single microcrack situated at the point 
$\bY_\pm$ (derived in \cite{roaz-ijss2011}) 
\beq
\label{mc}
\frac{\Delta K_\modIII}{K_\modIII^{(0)}} \sim \frac{1}{2} \frac{l^2}{d^2} \frac{\mu_\mp}{\mu_+ + \mu_-} 
\cos\left(\frac{3\varphi}{2} - \alpha\right)\cos\left(\frac{\varphi}{2} - \alpha\right), \quad a \to \infty.
\eeq
By summation over all the defects and assuming that the materials are identical $\mu_+ = \mu_-$, we obtain as $a \to \infty$
\beq
\label{sum1}
\barr{r}
\ds
\phi \sim \frac{al^2}{2h^2} \left\{ 
\sum_{j = 1}^{N^+} 
\frac{\frac{h^2}{j^2w^2}}{\left(1 + \frac{h^2}{j^2w^2}\right)^2}\left( \sqrt{1 + \frac{h^2}{j^2w^2}} + 2 \frac{h}{jw} \sin 2\alpha \right) \right. \\[6mm]
\ds
\left. -\sum_{j = 1}^{N^-} 
\frac{\frac{h^2}{j^2w^2}}{\left(1 + \frac{h^2}{j^2w^2}\right)^2}\left( \sqrt{1 + \frac{h^2}{j^2w^2}} + 2 \frac{h}{jw} \sin 2\alpha \right)
\right\}.
\earr
\eeq
It is noted that the contribution of the defects behind the crack tip, second sum in (\ref{sum1}), equals, but with the opposite sign, the contribution of defects 
ahead of the crack tip, first sum. We deduce that the incremental crack advance $\phi$ becomes zero when the crack tip reaches the centre of the array, as confirmed 
by the numerical results in Fig.\ \ref{figchA2B2n1w1=100}a.

For the case of Fig.\ \ref{figchA2B2n1w1=100}b, rigid line inclusions above and microcrack below the interface, we use again the formula (\ref{mc}) and also the formula 
valid for a single rigid line inclusion situated at the point $\bY_\pm$ (derived in \cite{roaz-ijss2011}) 
\beq
\label{rl}
\frac{\Delta K_\modIII}{K_\modIII^{(0)}} \sim -\frac{1}{2} \frac{l^2}{d^2} \frac{\mu_\mp}{\mu_+ + \mu_-} 
\sin\left(\frac{3\varphi}{2} - \alpha\right)\sin\left(\frac{\varphi}{2} - \alpha\right), \quad a \to \infty,
\eeq
thus obtaining, for identical materials, as $a \to \infty$
\beq
\label{sum2}
\phi \sim \frac{al^2 \cos 2\alpha}{2h^2} \left\{-1 + 
\sum_{j = 1}^{N^+} \frac{\left(1 - \frac{h^2}{j^2w^2}\right)\frac{h^2}{j^2w^2}}{\left(1 + \frac{h^2}{j^2w^2}\right)^2} +
\sum_{j = 1}^{N^-} \frac{\left(1 - \frac{h^2}{j^2w^2}\right)\frac{h^2}{j^2w^2}}{\left(1 + \frac{h^2}{j^2w^2}\right)^2} \right\}.
\eeq
In this case, the contribution of the defects behind the crack tip comes with the same sign as the contribution of the defects ahead of the crack tip, which explains 
the perfect symmetry about the centre of the array observed in Fig.\ \ref{figchA2B2n1w1=100}b. Moreover, since $\cos 2\alpha$ is a common factor in (\ref{sum2}), 
this formula also explains the constant thickness of the amplification region, $\pi/4 < \alpha < 3\pi/4$, in Fig.\ \ref{figchA2B2n1w1=100}b.

Furthermore, it is possible to extend the analysis to the case of an infinite number of defects ahead of the crack tip, $N^+ \to \infty$, as the corresponding series 
in (\ref{sum1}) and (\ref{sum2}) are convergent. In the latter case, we obtain an explicit formula as $a \to \infty$
\beq
\label{sum2b}
\phi \sim \frac{al^2 \cos 2\alpha}{2h^2} \left\{-\frac{1}{2} - \frac{\left(\frac{\pi h}{w}\right)^2}{2 \sinh^2 \left(\frac{\pi h}{w}\right)} + 
\sum_{j = 1}^{N^-} \frac{\left(1 - \frac{h^2}{j^2w^2}\right)\frac{h^2}{j^2w^2}}{\left(1 + \frac{h^2}{j^2w^2}\right)^2} \right\}.
\eeq

Finally, we would like to discuss the effects of inhomogeneity on the crack propagation. For a bimaterial plane and microcracks arranged as shown in 
Fig.\ \ref{figchA2B2n1}a, results (not reported here) show that the shielding-amplification diagram and the propagation curves are very similar to the case of a 
homogeneous plane, reported in Fig.\ \ref{figchA2B2n1}a. This result is expected, since the loading is symmetrical and the defects configuration is also symmetrical 
(apart from the inclination of defects, which is symmetrical only for the angle $\alpha = 3\pi/4$).

\begin{figure}[!htcb]
\begin{center}
\includegraphics[width=6cm]{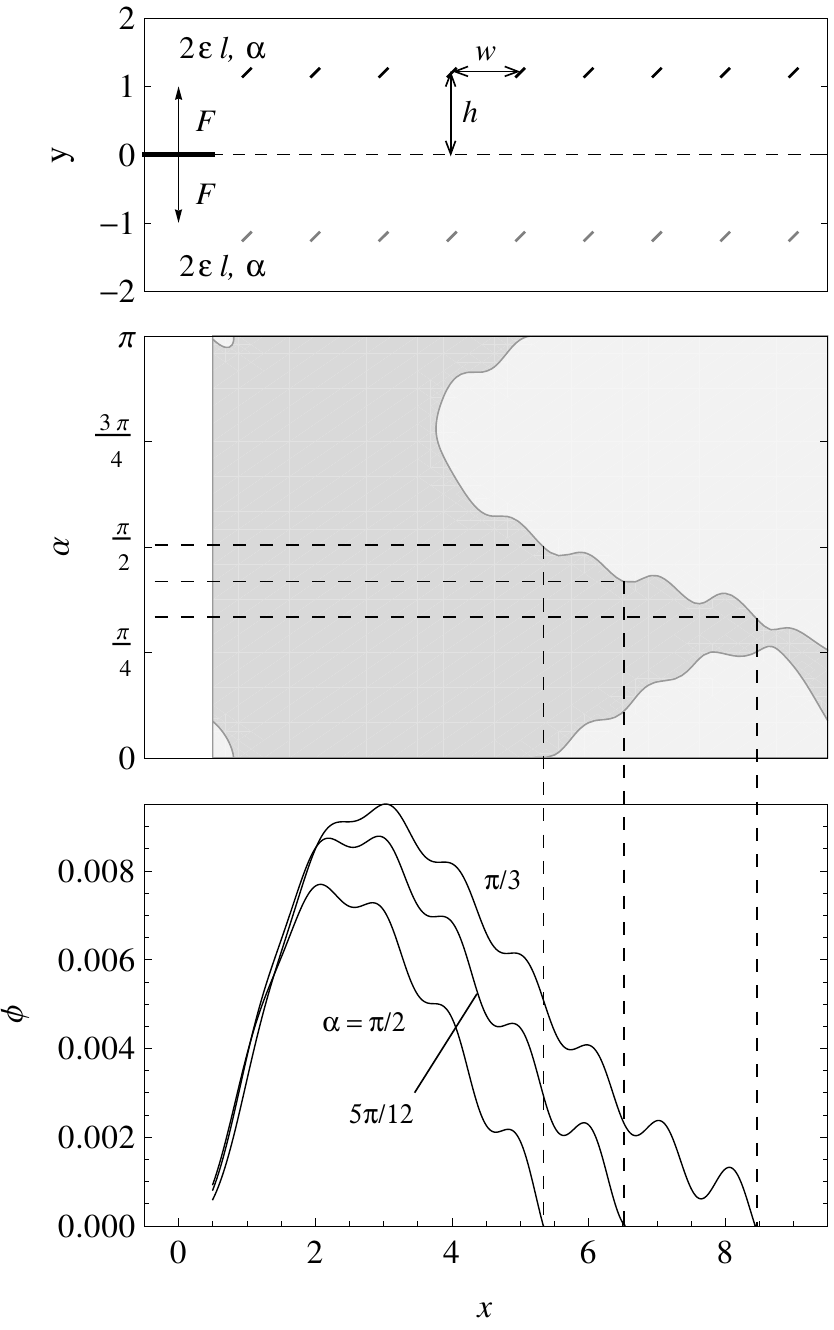}
\caption{\footnotesize Effects of the inhomogeneity. The figure corresponds to the case shown in Fig.\ \ref{figchA2B2n1}b but for a bimaterial plane with contrast 
parameter $\eta = -0.67$.} 
\label{figchB2n1in}
\end{center}
\end{figure}

We consider now the case of rigid line inclusions above and microcracks below the interface, as shown in Fig.\ \ref{figchA2B2n1}b, however now for a bimaterial plane with 
contrast parameter $\eta = -0.67$. In this case results, shown in Fig.\ \ref{figchB2n1in}, are completely different from the case of a homogeneous plane, compare with 
Fig.\ \ref{figchA2B2n1}b. Now the shielding-amplification diagram shows dependence on the angle $\alpha$ of orientation of the defects with respect to the $x_1$-axis. 
Consequently, the total crack elongation at arrest is different for different $\alpha$'s, as shown by the three propagation curves in the lower part of the figure.
Also, if we swap the two materials, such that $\eta = 0.67$, results (not reported here) show that the crack cannot propagate, since a shielding effect takes place 
instead of amplification effect at the initial position of the crack tip.
%%%%%%%%%%%%%%%%%%%%%%%%%%%%%%%%%%%%%%%%%%%%%%%%%%%%%%%%%%%%%%%%%%%%%%
%%%%%%%%%%%%%%%%%%%%%%%%%%%%%%%%%%%%%%%%%%%%%%%%%%%%%%%%%%%%%%%%%%%%%%

\section{Conclusions}
\label{secconc}

The problem of propagation of a main interfacial crack in a bimaterial plane containing a finite channel of small line defects and subject to antiplane loading 
conditions is analysed on the basis of asymptotic formulae derived by the authors. Under the assumption of quasi-static propagation, the shielding-amplification 
effect of the microdefects on the main crack tip is obtained at each position of the crack tip and the incremental crack advance is derived asymptotically.

Two numerical examples are shown for two different microdefect arrangements. Results show that the channel of microdefects can facilitate (amplification) or prevent 
(shielding) the propagation of the main crack. In particular, the orientation of the microdefect can be tuned such that the main crack stops at a given position along 
the channel. We discuss the possibility to extend the asymptotic formulation to the case of a channel with infinite number of small defects.

\vspace{5mm}
{\bf Acknowledgements.} This research was supported by the Research-In-Groups (RiGs) programme of the International Centre for Mathematical 
Sciences, Edinburgh, Scotland. In addition, A.P. and G.M. gratefully acknowledge the support from the European Union Seventh Framework Programme
under contract numbers PIEF-GA-2009-252857 and PIAP-GA-2009-251475, respectively.

\bibliographystyle{abbrv}
\bibliography{biblio}

\end{document}